\documentclass[pre]{revtex4}
\usepackage{amsmath}
\usepackage{graphicx}
\begin{document}
\title{Solitary wave state in the nonlinear Kramers equation for self-propelled particles}
\author{Hidetsugu Sakaguchi and Kazuya Ishibashi}
\affiliation{Department of Applied Science for Electronics and Materials,
Interdisciplinary Graduate School of Engineering Sciences, Kyushu
University, Kasuga, Fukuoka 816-8580, Japan}
\begin{abstract}
We study collective phenomena of self-propagating particles using the nonlinear Kramers equation. A solitary wave state appears from an instability of the spatially uniform ordered state with nonzero average velocity. Two solitary waves with different heights merge into a larger solitary wave. An approximate solution of the solitary wave is constructed using a self-consistent method. The phase transition to the solitary wave state is either first-order or second-order, depending on the control parameters.   
\end{abstract}
\maketitle
\section{Introduction}
The collective motion of self-propelled particles has been intensively studied since Vicsek et al. proposed a simple dynamical model for the collective phenomenon.~\cite{Vicsek}. A large population of self-propelled units is called active matters. Typical active matters are biological systems such as schools of fish, flocks of birds, and bacterial colonies; however, there are also various nonbiological active matters such as vibrating granular rods and Janus particles. Various agent-based models  similar to the Vicsek model have been proposed and numerically studied to investigate the dynamical behaviors of a large population of self-propelled particles~\cite{Vicsek2}. Besides these studies of collective motion, there are many studies on the dynamics of each self-propelled particle.~\cite{Ohta} Theoretically, phenomenological hydrodynamic equations for active matters have been investigated~\cite{Toner, Marchetti}.  Collective directional motion appears as a kind of order-disorder transition in a large population of self-propelled particles. The long-range order exists even in two dimensions, in contrast to thermal equilibrium systems of short-range interactions at a finite temperature, where the long-range order does not appear owing to the Mermin--Wagner theorem. The transition to the collective motion is considered to be a first-order one, that is, the order parameter jumps at the transition~\cite{gre}. 

Many self-propelling animals make a flock. That is, there are localized clusters of self-propelled units. There might be many mechanisms of clustering~\cite{Shimoyama}, but it was found even in the Vicsek model that localized regions of high density propagate with a constant velocity similarly to solitary waves.~\cite{Chate} The solitary wave state appears near the transition point.  This phenomenon was investigated as the instability of the spatially uniform state in hydrodynamic equations for active matters.~\cite{Bertin,Mishra,Gopinath} Ihle studied the solitary wave state by performing numerical simulations of a model system based on Enskog-like kinetic theory~\cite{Ihle}. 
However, the mechanism of the formation is not fully understood. Since the hydrodynamic system is  dissipative, the solitary wave is related to dissipative solitons found in many nonlinear-nonequilibrium systems~\cite{Thual, Sakaguchi2, Krischer, Descalzi}. 

On the other hand, the original Vicsek model has a form similar to nonlocally coupled oscillators~\cite{Kuramoto}. The nonlocally coupled stochastic oscillators can be studied with the Fokker--Planck equation using the mean field approximation~\cite{Kuramoto2,Sakaguchi}.  Here, we  apply the method of the nonlinear Fokker--Planck equation to the problem of self-propelled particles. The dynamics of the probability distributions for the velocity and position of self-propelled particles is studied using the nonlinear Kramers equation. We show that a solitary wave state appears owing to an instability from a spatially uniform ordered state in the nonlinear Kramers equation. The solitary wave state is a nonequilibrium state which is generated from the nonequilibrium phase transition of self-propelled particles. The nonequilibrium phase transition can be clearly seen, because the probability distributions are directly obtained in the numerical simulation of the nonlinear Kramers equation.  

\section{Nonlinear Kramers Equation}
We consider a large population of nonlocally coupled self-propelled particles in one dimension for the sake of simplicity.   
The model equation is expressed with the Langevin equation
\begin{subequations}
\begin{align}
\frac{dx_i}{dt}&=v_i,\\
\frac{dv_i}{dt}&=\mu v_i-v_i^3+g\sum_{j=1}^Ne^{-\alpha\{1-\cos(2\pi(x_j-x_i)/L)\}}(v_j -v_i)+\xi_i(t), 
\end{align}
\end{subequations}
where $x_i$ and $v_i$ are the $x$ coordinate and velocity of the $i$th element, $L$ is the system size, and $\xi_i(t)$ is Gaussian white noise 
satisfying $\langle \xi_i(t)\xi_j(t')\rangle=2T\delta_{i,j}\delta(t-t')$. If $g=T=0$, each elemental particle becomes a self-propelled particle of velocity $v=\pm \sqrt{\mu}$ for $\mu>0$. The nonlocal interaction in the spatially periodic one-dimensional system of size $L$ is expressed by the third summation term on the right-hand side of Eq.~(1b). The nonlocal interaction tends to make the velocities of active particles uniform. Similar terms to align the velocities are assumed in the Vicsek model and many other related models. However, our model is simpler than those models in that it is a one-dimensional model in contrast to the Vicsek type models in two dimensions. The integral kernel $e^{-\alpha\{1-\cos(2\pi (x^{\prime}-x)/L)\}}$ is approximated by the Gaussian function $e^{-(2\pi^2\alpha/L^2)(x^{\prime}-x)^2}$ if $x^{\prime}-x$ is sufficiently small. 
For example, the Gaussian approximation is good for $\alpha=10$ and $L=5$, which are typical parameter values used in later numerical simulations. 
We chose the integral kernel because the relatively simple function is well approximated locally by the Gaussian and  has spatial periodicity of period $L$.

The Kramers equation corresponding to the Langevin equation is expressed as
\begin{equation}
\frac{\partial P}{\partial t}=-\frac{\partial}{\partial x}\left (vP\right )-\frac{\partial}{\partial v}\left  [\left \{\mu v-v^3+g\int_0^Le^{-\alpha\{1-\cos(2\pi (x^{\prime}-x)/L)\}}\rho(x^{\prime})(u(x^{\prime}) -v)dx^{\prime}\right \}P\right ]+T\frac{\partial^2P}{\partial v^2} \label{kr},
\end{equation} 
where $P(x,v,t)$ is the probability density function, $\rho(x)=\int_{-\infty}^{\infty}P(x,v,t)dv$ denotes the density at  position $x$, and $u(x)=\int_{-\infty}^{\infty}vP(x,v,t)dv/\rho(x)$ denotes the average velocity at the position. 
  If each particle interacts nonlocally with a sufficiently large number of particles,  a mean-field approximation can be applied even in one dimension~\cite{Kuramoto2,Sakaguchi}.   
  In our model of nonlocally coupled active particles, we assume a kind of mean-field approximation by replacing the summation in Eq.~(1b) with the integral expressed in terms of the density $\rho(x)$ and average velocity $u(x)$. 
The Kramers equation is a deterministic equation for nonlocally coupled self-propelled particles, and thus phase transitions can be treated as bifurcations in the nonlinear equation. The Kramers equation is a nonlinear equation, because $\rho(x)$ and $u(x)$ in the second term on the right-hand side of Eq.~(2) are determined by $P(x,v,t)$. 
Furthermore, the normalization condition $\int_0^L\int_{-\infty}^{\infty}P(x,v,t)dvdx=1$ is assumed in this paper. 

If the system is spatially uniform, $\partial P/\partial x=0$ and $\rho(x)=1/L$. Then, Eq.~(2) is reduced to  
\begin{equation}
\frac{\partial P}{\partial t}=-\frac{\partial}{\partial v}\left  [\left \{\mu v-v^3+\frac{g}{L}\int_0^Le^{-\alpha\{1-\cos(2\pi (x^{\prime}-x)/L)\}}(u(x^{\prime}) -v)dx^{\prime}\right \}P\right ]+T\frac{\partial^2P}{\partial v^2}.
\end{equation} 
The free energy density $F/L$ is defined as
\begin{eqnarray}
F/L&=&\int_{-\infty}^{\infty}\left  [\left \{-(1/2)\mu v^2+(1/4)v^4+\frac{g}{L}\int_0^Le^{-\alpha\{1-\cos(2\pi (x^{\prime}-x)/L)\}}(-u(x^{\prime}) v +(1/2)v^2)dx^{\prime}\right \}P\right ]dv\nonumber\\
& &+T\int_{-\infty}^{\infty}P\log Pdv.
\end{eqnarray} 
The free energy decreases in the time evolution of Eq.~(3), that is, $F$ is the Lyapunov function of Eq.~(3).

\begin{figure}[tbp]
\begin{center}
\includegraphics[height=3.5cm]{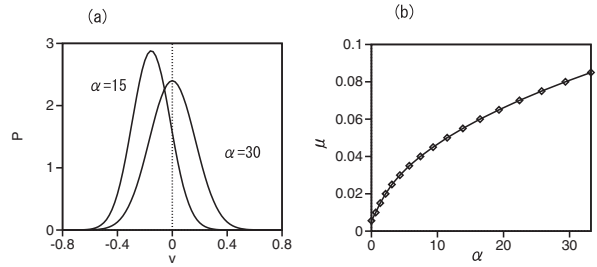}
\end{center}
\caption{(a) Equilibrium distributions at $\alpha=15$ and 30 for $\mu=0.08$, $T=0.1$, $g=50$, and $L=5$. (b) Critical line in the parameter space $(\alpha,\mu)$ for $\mu=0.08$, $T=0.1$, $g=50$, and $L=5$.}
\label{fig1}
\end{figure}
The equilibrium probability distribution is given by 
\begin{equation}
P_{eq}(v)=ce^{[(1/2)\mu v^2-(1/4)v^4-r\{-uv+(1/2)v^2\}]/T},
\end{equation}
where $r=g/L\int_0^Le^{-\alpha\{1-\cos(2\pi (x^{\prime}-x)/L)\}}dx$ and $c=1/\int_{-\infty}^{\infty}P_{eq}(v)dv$ is the normalization constant. The spatially uniform average velocity $u$ satisfies
\begin{equation}
u=\int_{-\infty}^{\infty}P_{eq}(v)vdv,
\end{equation}
where $P_{eq}$ includes $u$ and therefore Eq.~(6) is a self-consistent equation. The average velocity $u$ is a solution to this self-consistent equation.   
If the Taylor series expansion is applied to the right-hand side of Eq.~(6) with respect to $u$, only the odd-power terms of $u$ appear, that is, the Taylor expansion is expressed as $c_1u+c_3u^3+c_5u^5+\cdots$, because the function $e^{[(1/2)\mu v^2-(1/4)v^4-(1/2)rv^2]/T}$  in $P_{eq}(v)$ is an even function of $v$ and $e^{ruv}$ is expanded as $1+ruv+(1/2)(ruv)^2+\cdots$.  A pitchfork bifurcation can occur at a certain parameter value. 

Figure 1(a) shows  equilibrium distributions $P_{eq}(v) $ at $\alpha=15$ and 30 for $\mu=0.08$, $T=0.1$, $g=50$, and $L=5$. The average value $u$ can be obtained by solving the self-consistent equations Eqs.~(5) and (6) with the iteration method. The average velocity $u$ is 0 at $\alpha=30$ and takes a nonzero value at $\alpha=15$ as a result of the symmetry-breaking phase transition.  In Fig.~1(a), the average value $u$ is negative but it can take a positive value owing to the symmetry.  There is an order-disorder phase transition at $\alpha\simeq 29.4$. This transition is a second-order phase transition, and the absolute value of the average velocity increases continuously from 0.  
The first coefficient $c_1$ of the Taylor expansion becomes 1 at the phase transition. This condition is explicitly expressed as 
\begin{equation}
\frac{\int_{-\infty}^{\infty}(r/T)v^2e^{(\mu/2)v^2-(1/4)v^4/4-(r/2)v^2}dv}{\int_{-\infty}^{\infty}e^{(\mu/2)v^2-(1/4)v^4/4-(r/2)v^2}dv}=1.
\end{equation}
Figure 1(b) shows the numerically obtained critical line satisfying Eq.~(7) in the parameter space $(\alpha,\mu)$. For $\mu>0$, each particle is a self-propelled particle, but collective motion appears above the critical line owing to the effect of noises.

\section{Solitary Wave State in the Nonlinear Kramers Equation}
The spatially uniform state is not always stable. 
 A traveling solitary wave appears when the spatially uniform self-driving state becomes unstable.  We performed a direct numerical simulation of Eq.~(2) with the simple Euler method of timestep $\Delta t=0.0005$. The coordinate $x$ and velocity $v$ are respectively discretized by $\Delta x=1/40$ and $\Delta v=1/50$.  

Figure 2(a) shows the time evolution of $\rho(x,t)=\int_{-\infty}^{\infty}P(x,v,t)dv$ for $\mu=0.2$, $\alpha=180$, $T=0.1$, $g=50$, and $L=15$. The solitary wave propagates with a constant velocity $v_0\simeq -0.4116$. The peak density is $\rho(x)\simeq 0.588$ and the density takes a small value of $\rho(x)\simeq 0.0144$ in the region distant from the peak position. Even when initial conditions were changed, we did not obtain a pulse train state as a final stationary state. Figure 2(b) shows the time evolution from an initial condition with two peaks of $\rho(x)$ for the same parameter values as in Fig.~2(a). The solitary wave with the higher peak moves faster than that with the lower peak. The higher solitary wave catches up with the lower one and the two solitary waves merge into an even higher solitary wave. Figure 2(c) shows the time evolution of the head-on collision of two solitary waves moving in opposite directions with slightly different amplitudes at the same parameters as in Fig.~2(a).  The two solitary waves merge and one solitary wave traveling in the right direction arises, which is the direction of the solitary wave with larger amplitude. Generally,  various pheneomena such as pair annihilation, interpenetration, and the formation of a bound state occur at the head-on collision of two dissipative solitons, depending on the control parameters~\cite{Krischer, Descalzi}. The merging in our system corresponds to the formation of a bound state, which is one of the typical behaviors. Ihle found the interpenetration of two solitary waves at multiple head-on collisions in a numerical simulation of the model based on kinetic theory. However, the amplitude difference between the two solitary waves increases at each collision, and finally only one solitary wave survives.  

Figure 2(d) shows the relation between $u(x)$ and $1/\rho(x)$ for the final state in Fig.~2(a). The average velocity $u(x)$ becomes 0 in the lowest-density region of $\rho\simeq 0.0144$, because the disordered state appears when $\rho<\rho_c=0.062$ in the spatially uniform state.  
Figure 2(d) shows clearly that  $\rho(x,t)$ and $u(x,t)$ satisfy $\rho(x)\propto 1/(u(x)-v_0)$.
\begin{figure}
\begin{center}
\includegraphics[height=3.5cm]{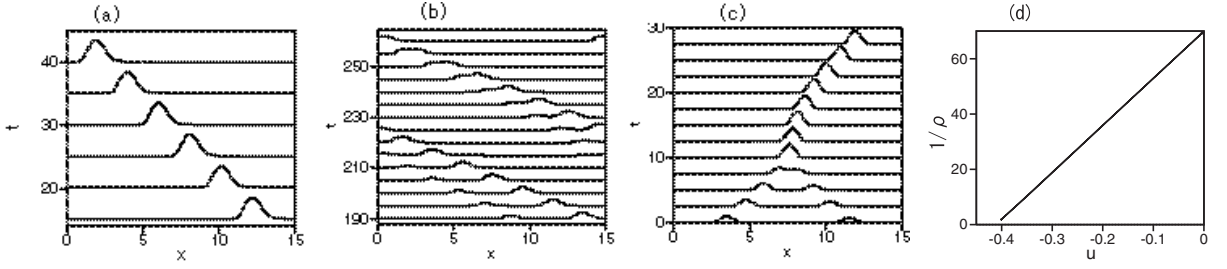}
\end{center}
\caption{(a) Traveling solitary wave state at $\mu=0.2$, $\alpha=180$, $T=0.1$, $g=50$, and $L=15$. (b) Merging process of two solitary waves at $\mu=0.2$, $\alpha=180$, $T=0.1$, $g=50$, and $L=15$. (c) Head-on collision of two solitary waves at $\mu=0.2$, $\alpha=180$, $T=0.1$, $g=50$, and $L=15$. (d) Relationship between the average velocity $u(x)$ and $1/\rho(x)$.    
}
\label{fig2}
\end{figure}
\begin{figure}[tbp]
\begin{center}
\includegraphics[height=3.5cm]{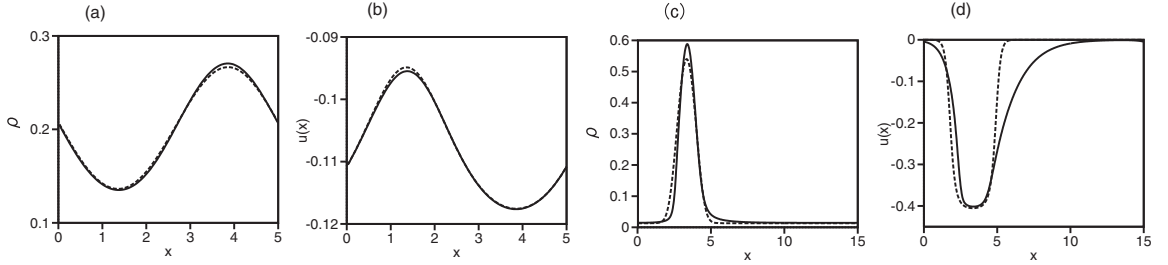}
\end{center}
\caption{(a) Comparison of the density $\rho(x)$ (solid line) obtained by direct numerical simulation and $\rho(x)$ (dashed line) obtained by the self-consistent method for $\mu=0.07$, $\alpha=16$, $T=0.1$, $g=50$, and $L=5$. (b) Comparison of the average velocity $u(x)$ (solid line) obtained by direct numerical simulation and the average velocity $u(x)$ (dashed line) obtained by the self-consistent method for the same parameters as in (a). (c) Comparison of the densities $\rho(x)$ obtained by the two methods for $\mu=0.2$, $\alpha=180$, $T=0.1$, $g=50$, and $L=15$. (d) Comparison of the average velocities $u(x)$ obtained by the two methods for the same parameters as in (c).
}
\label{fig3}
\end{figure}
 This nontrivial relation is derived as follows. The integration of Eq.~(2) with respect to $v$ yields
\begin{equation}
\frac{\partial \rho}{\partial t}=-\frac{\partial}{\partial x}\left (u\rho\right )
\end{equation} 
if $P(x, v,t)\rightarrow 0$ and $\partial P/\partial v\rightarrow 0$  are satisfied for $v\rightarrow \pm \infty$. If the wave propagates with a constant velocity $v_0$, then $\partial\rho/\partial t=-v_0\partial \rho/\partial x$, therefore,
\begin{equation}
\frac{\partial}{\partial x}\left\{ (u(x)-v_0)\rho\right \}=0.
\end{equation}
Thus, the relation $\rho(x,t)\propto 1/(u(x)-v_0)$ is satisfied. 

The explicit form of the solitary wave is not yet known. We assume an approximate solution of the form $P(x,v,t)=\rho(x-v_0t)f(v,t)$ to understand the solitary wave state qualitatively. That is, $P(x,v,t)$ is expressed as the product of the propagating density and the probability distribution of $v$ under the nonuniform density $\rho(x-v_0t)$. 
As an approximation, the probability distribution $f$ of $v$ satisfies the Fokker--Planck equation:
\begin{equation}
\frac{\partial f}{\partial t}=-\frac{\partial}{\partial v}\left  [\left \{\mu v-v^3+g\int_0^Le^{-\alpha\{1-\cos(2\pi (x^{\prime}-x)/L)\}}\rho(x^{\prime})(u(x^{\prime}) -v)dx^{\prime}\right \}f\right ]+T\frac{\partial^2f}{\partial v^2}.
\end{equation} 
The equilibrium distribution $f$ is given by
\begin{equation}
f_{eq}(x,v)\propto e^{\{(1/2)\mu v^2-(1/4)v^4+Av-(1/2)Bv^2)\}/T} \label{veq},
\end{equation}
where 
\[A(x)=g\int_0^Le^{\alpha\{1-\cos(2\pi (x^{\prime}-x)/L)\}}\rho(x^{\prime})u(x^{\prime})dx^{\prime},\;B(x)=g\int_0^Le^{\alpha\{1-\cos(2\pi (x^{\prime}-x)/L)\}}\rho(x^{\prime})dx^{\prime}.\] 
$A(x)$ and $B(x)$ are spatially dependent  mean fields which determine the velocity distribution of self-propelled particles. The average velocity $u(x)$ is given by $u(x)=\int_{-\infty}^{\infty}vf_{eq}(v)dv$. The spatially dependent average velocity $u(x)$ is determined self-consistently because $A(x)$ in $f_{eq}(v)$ is calculated using $u(x)$. The quantity $u(x)$ is numerically obtained by iterating the integrals $A(x)=g\int_0^Le^{\alpha\{1-\cos(2\pi (x^{\prime}-x)/L)\}}\rho(x^{\prime})u(x^{\prime})dx^{\prime}$ and $u(x)=\int_{-\infty}^{\infty}vf_{eq}(v)dv$.   
If the average velocity $u(x)$ is obtained, the average density $\rho(x)$ is evaluated as $\rho(x)\propto 1/|u(x)-v_0|$. 
We attempted to obtain the solutions of $\rho(x)$, $u(x)$, $A(x)$, and $B(x)$ by this iteration method for several values of $v_0(<0)$. We found that $\rho(x)$ increases to infinity for $v_0>v_{0c}$ and $\rho(x)$ decreases to 0 for $v_0<v_{0c}$.  There is  a critical value of $v_{0c}$ at which self-consistent solutions $\rho(x)$ and $u(x)$ are obtained.  

Figure 3(a) shows the density $\rho(x)$ (solid line) obtained by direct numerical simulation and $\rho(x)$ (dashed line) obtained by the self-consistent method at $v_0=-0.14127$ for $\mu=0.07$, $\alpha=16$, $T=0.1$, $g=50$, and $L=5$. 
Figure 3(b) compares the average velocity $u(x)$ (solid line) obtained by direct numerical simulation with $u(x)$ (dashed line) obtained by the self-consistent method. 
The parameters are close to the instability point of the spatially uniform state and a sinusoidal wave appears. 
Fairly good agreement is seen between the numerical results and the results obtained by the self-consistent method. The critical velocity $v_{0c}=-0.14127$ is also close to the propagating velocity $v_0=-0.1397$ of the solitary wave in the direct numerical simulation. 
Figures 3(c) and 3(d) show the densities $\rho(x)$ and average velocities $u(x)$ obtained by the direct numerical simulation and the self-consistent method at $v_0=-0.4154$ for $\mu=0.2$, $\alpha=180$, $T=0.1$, $g=50$, and $L=15$. 
These parameters correspond to those used in Fig.~2. 
A solitary wave state is obtained by the self-consistent method. The numerically obtained velocity $v_0=-0.4116$ of the solitary wave is close to the critical value $v_{0c}=-0.4154$. However, some disagreement is seen, probably because the solitary wave state appears far from the instability point, where the approximation $P(x,v,t)=\rho(x-v_0t)f_{eq}(v)$ becomes less accurate.     

\begin{figure}
\begin{center}
\includegraphics[height=3.5cm]{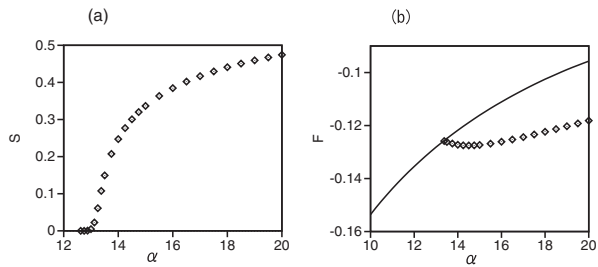}
\end{center}
\caption{(a) Fourier amplitude $S$ as a function of $\alpha$ for $\mu=0.08$, $T=0.1$, $g=50$, and $L=5$. (b) Free energy $F$ as a function of $\alpha$ for $\mu=0.08$. The solid line corresponds to the spatially uniform state and the rhombi denote the solitary wave states obtained by direct numerical simulation of Eq.~(2). 
}
\label{fig4}
\end{figure}
\begin{figure}
\begin{center}
\includegraphics[height=3.5cm]{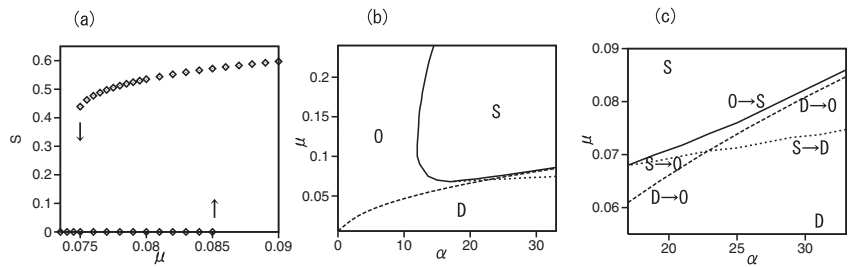}
\end{center}
\caption{(a) Fourier amplitude $S$ as a function of $\mu$ for $\alpha=33$, $T=0.1$, $g=50$, and $L=5$. (b) Phase diagram in the parameter space $(\alpha,\mu)$ for $g=50$ and $L=5$. `S': solitary wave state, `D': disordered state, and `O': spatially uniform ordered state. (c) Magnification of (b). The transition lines $S\rightarrow D$, $D\rightarrow 0$, $0\rightarrow S$, and $S\rightarrow 0$ are shown. 
}
\label{fig5}
\end{figure}

Phase transitions occur in the nonlinear Kramers equation when the parameters are changed. To characterize the solitary wave state, we use the Fourier amplitude $S$  of wavenumber $2\pi/L$ defined as
\[S=\left [\left\{\int_0^L\int_{-\infty}^{\infty}P(x,v)\sin(2\pi x/L)dvdx\right \}^2+\left \{\int_0^L\int_{-\infty}^{\infty}P(x,v)\cos(2\pi x/L)dvdx\right \}^2\right ]^{1/2}.\]
Figure 4(a) shows $S$ as a function of $\alpha$ for  $\mu=0.08$, $T=0.1$, $g=50$, and $L=5$. $S$ increases continuously from 0 at $\alpha\simeq 13$. 
The transition from the spatially uniform ordered state to the solitary wave state is a continuous transition.  Figure 4(b) shows $F$ defined by
\begin{eqnarray}
F&=&\int_0^L\int_{-\infty}^{\infty}\left  [\left \{-\frac{\mu v^2}{2}+\frac{v^4}{4}+g\int_0^Le^{-\alpha\{1-\cos(2\pi (x^{\prime}-x)/L)\}}\rho(x^{\prime})(-u(x^{\prime}) v +v^2/2)dx^{\prime}\right \}P\right ]dvdx\nonumber\\
& &+T\int_0^L\int_{-\infty}^{\infty}P\log Pdvdx. \label{free}
\end{eqnarray}
$F$ is  a simple extension of the free energy given by Eq.~(4) for the spatially uniform state to the spatially nonuniform state. We consider $F$ as the free energy for Eq.~(2) in this paper, although $F$ is not exactly the Lyapunov function of Eq.~(2). (It is generally difficult to find the Lyapunov function in a nonequilibrium system.) The solid line denotes $F$ for the spatially uniform state obtained using Eqs.~(5) and (6). The rhombi denote $F$ for the solitary wave state obtained by the direct numerical simulation of Eq.~(2). The solitary wave state has lower free energy $F$ than the spatially uniform state owing to the lower energy in the first term of Eq.~(\ref{free}). It is interesting that the self-propelled particles gather together, even if mutual attraction is not assumed.        

 Figure 5(a) shows $S$ as a function of $\mu$ for $\alpha=33$, $T=0.1$, $g=50$, and $L=5$. The solitary wave state jumps to the disordered state for $\mu<0.0745$. The disordered state changes into the spatially ordered state at $\mu=0.0835$ and the spatially ordered state jumps to the solitary wave state at $\mu=0.086$. There is a small bistable region of the disordered state and the solitary wave state. 
Figure 5(b) shows a phase diagram in the parameter space $(\alpha,\mu)$ for $g=50$ and $L=5$. The solid line shows the instability line of the spatially uniform ordered state. The dashed line is the phase transition line from the disordered state to the spatially uniform ordered state shown in Fig.~1. 
The dotted line denotes the transition line from the solitary wave state to the spatially uniform states. Figure 5(c) is a magnification of Fig.~5(b) around the parameter region where the three lines cross.   
The transition between the spatially uniform states and the solitary wave state  is a first-order transition for $\alpha>17$. That is, when $\mu$ is increased from a sufficiently small value for $\alpha>17$, the disordered state `D' changes into the ordered state `O' at the transition line D$\rightarrow$ O in Fig.~5(c) and then the ordered state `O' changes into the solitary wave state at O$\rightarrow$ S. On the other hand, when $\mu$ is decreased from a sufficiently large value, the solitary wave state changes into the disordered state at the line S$\rightarrow$ D for $\alpha>23$ and the solitary wave state changes into the ordered state at the line S$\rightarrow$ O for $17<\alpha<23$.    
The transition, however, becomes second-order for $\alpha<17$. That is, the disordered state changes into the ordered state, and the ordered state continuously changes into the solitary wave state as shown in Fig.~4(a). 
\section{Summary}
We have investigated the nonlinear Kramers equation for one-dimensional self-propelled particles. This equation is based on the mean-field approximation, which can be used in nonlocally interacting systems. 
Several problems regarding the collective motion of self-propelled particles are clarified by the deterministic nonlinear Kramers equation.  
A solitary wave state appears as a stable state in the nonlinear Kramers equation. The phase transition to the solitary wave state occurs even in one dimension, because our model is a nonlocally coupled system and the mean-field approximation can be applied. Furthermore, we have found that the average concentration $\rho(x)$ and average velocity $u(x)$ satisfy a simple relation, and an approximate solution of the solitary wave state is constructed using the self-consistent method.  The transition from the disordered state to the solitary wave state becomes discontinuous, depending on the control parameters.

We consider that the solitary wave state in our one-dimensional model corresponds to the traveling band found in Vicsek's two-dimensional model, because both states are solitary waves and appear near the transition point to the collective motion. In the head-on collision of two solitary waves, we found that only one solitary wave survives after the collision. This is slightly different from the result of Ihle's numerical simulation that two solitary waves interpenetrate with each other at a head-on collision. However, only one solitary wave survives in the final state of Ihle's simulation, which is similar to our numerical result. 
 
We have found that the solitary wave state appears naturally in the direct numerical simulation of the nonlinear Kramers equation; however, the mechanism of formation of the solitary wave state is still not clarified. It is interesting that  self-propelled particles gather together in the solitary wave state even though mutual attraction is absent. We have considered a qualitative mechanism of the formation of the solitary wave state using the free energy given by Eq.~(\ref{free}), although  more detailed theoretical analysis is left to future study. Furthermore, we would like to investigate our model in wide parameter ranges in the future since we have studied the model equation only for limited parameter sets.

\end{document}